\documentclass[12pt]{article}
\usepackage{times}
\usepackage{amsmath,amssymb,mathrsfs}
\usepackage{epsfig,graphics}
\usepackage{color}
\begin{document}
\hsize = 6. in
\vsize = 9.0 in
\hoffset = -0.5 in
\voffset = -0.85 in
\baselineskip = 0.23 in

\def\rd{{\rm d}}
\def\vx{{\bf x}}
\def\vz{{\bf z}}
\def\vi{{\bf i}}
\def\vj{{\bf j}}
\def\vk{{\bf k}}
\def\vdelta{\mbox{\boldmath$\delta$}}
\def\mD{{\bf D}}
\def\mI{{\bf I}}
\def\mJ{{\bf J}}
\def\vf{{\bf f}}
\def\tE{\tilde{E}}
\def\vB{{\bf B}}
\def\vX{{\bf X}}
\def\vY{{\bf Y}}
\def\vZ{{\bf Z}}

\title{Universal Ideal Behavior and Macroscopic Work Relation of
Linear Irreversible Stochastic Thermodynamics
}

\author{Yi-An Ma\footnote{Email: yianma@u.washington.edu }\hspace{0.2cm}
  and Hong Qian\footnote{
Email: hqian@u.washington.edu}\\[10pt]
Department of Applied Mathematics\\
University of Washington, Seattle\\
WA 98195-3925, U.S.A.}

\maketitle

\begin{abstract}
We revisit the Ornstein-Uhlenbeck (OU) process as the fundamental
mathematical description of linear irreversible phenomena, with
fluctuations, near an equilibrium. By identifying the underlying circulating dynamics in a stationary process as the natural generalization of classical conservative mechanics, a bridge between a family
of OU processes with equilibrium fluctuations and thermodynamics is established through the celebrated Helmholtz theorem. The Helmholtz theorem provides an emergent macroscopic ``equation of state'' of the entire system, which exhibits a universal ideal thermodynamic behavior.
Fluctuating macroscopic quantities are studied from the
stochastic thermodynamic point of view and a non-equilibrium
work relation is obtained in the macroscopic picture, which
may facilitate experimental study and application of the
equalities due to Jarzynski, Crooks, and Hatano and Sasa.

\end{abstract}

\tableofcontents

\section{Introduction}

	Gaussian fluctuation theory is one of the most successful
branches of equilibrium statistical mechanics \cite{landau,callen}.
Since the work of Onsager and Machlup \cite{onsager_53a,onsager_53b}, the Ornstein-Uhlenbeck
process (OUP) has become the stochastic, mathematical description
of dynamic, linear irreversible phenomena \cite{wax}.
It has been extensively discussed in the literature in the past \cite{cox,lax_1,lax_2,fox}.
Several recent papers studied particularly the OUP without detailed balance \cite{qian_prsa_01,kwon_ao_thouless_pnas,dotsenko-2013}.
In recent years, taking stochastic process rigorously developed
by Kolmogorov as the
mathematical representation, stochastic thermodynamics has emerged
as the finite-time thermodynamic theory of mesoscopic systems,
near and far from equilibrium \cite{seifert_rpp,ge-chinese-am,cvdb-me,altaner}.
The fundamental aspects of this new
development are the mathematical notion of stochastic entropy
production \cite{qqg,seifert_prl,ge_jiang}, novel thermodynamic
relationships collectively known as nonequilibrium work equalities,
and fluctuation theorems \cite{jarzynski,kurchan,crooks,lebowitz-spohn,sasa}, and the
mathematical concept of non-equilibrium steady-state \cite{jqq,zqq,gqq}.

	Fundamental to all these advances is the notion of {\em time reversal}.
Newtonian dynamic equation, in Hamiltonian form:
\begin{equation}
        \frac{\rd x_i}{\rd t} = \frac{\partial H(x_i,y_i)}{\partial y_i}, \   \
       \frac{\rd y_i}{\rd t} = -\frac{\partial H(x_i,y_i)}{\partial x_i},
\label{1}
\end{equation}
is a canonical example of dynamics with time-reversal symmetry \cite{lamb-roberts}:
Under transformation $(t,x_i,y_i)\longrightarrow (-t,x_i,-y_i)$,
Eq. (\ref{1}) is invariant. This invariance requires that
$H(x_i,y_i)=H(x_i,-y_i)$: $H$ is usually a function of $y_i^2$ and
terms like $\vec{B}\cdot\vec{y}$, where $\vec{B}$ changes sign
upon time reversal such as a magnetic field with a Lorentz force.
Adopting this definition to linear stochastic processes, one has a
novel definition for {\em time reversibility} that is distinctly different
from that of Kolmogorov's, as we shall show below.

	Consider the linear stochastic differential equation
\begin{equation}
                \rd \vX(t) = -M(\alpha) \vX(t) \rd t  + \epsilon\Gamma\rd \vB_t,
\label{2}
\end{equation}
which is an OUP with parameters $\alpha$ and
$\epsilon$; $M$ and $\Gamma$ are two $n\times n$ constant
matrices, $\vB_t$ is standard Brownian motion.  We further
assume that all the eigenvalues of $M$ are strictly positive and
$\Gamma$ is non-singular.  According to the
concept of detailed balance, Eq. (\ref{2}) can be uniquely written as
\cite{vankampen,risken,aoping,qian_pla_2014}
\begin{subequations}
\begin{equation}
         \rd\vX(t) = -D\Big\{\Xi^{-1}
                     + \Big( D^{-1}M-\Xi^{-1}\Big) \Big\}\vX\rd t +\epsilon\Gamma \rd \vB_t,
\label{3a}
\end{equation}
where $D$ and $\Xi(\alpha)$ are positive definite matrices:
$D=\frac{1}{2}\Gamma\Gamma^T$ and $M\Xi+\Xi M^T=2D$.  If
one identifies the two terms inside $\{\cdots\}$ as dissipative (transient)
and conservative (perpetuate) motions, respectively, then a
time reversible process should be defined as a statistical
equivalence between the probability density of
a finite path $\{\vX(t_0)=x_0,\vX(t_1)=x_1,\cdots,\vX(t_n)=x_n\}$ in which $t_0<t_1\cdots<t_n$:
\[
                f\big(x_0,x_1,\cdots,x_n\big),
\]
and the probability density
\[
        f_{\vX^{\dag}(t_n)\vX^{\dag}(2t_n-t_{n-1})\cdots
               \vX^{\dag}(2t_n-t_0)}\big(x_n,x_{n-1},\cdots,
                     x_0 \big)
\]
in which the $\vX^{\dag}(t)$ follows the adjoint stochastic differential
equation \cite{qian_pla_2014,qian_epj_st}
\begin{equation}
         \rd\vX^{\dag}(t) = -D\Big\{\Xi^{-1}
                     -\Big( D^{-1}M-\Xi^{-1}\Big) \Big\}\vX^{\dag}\rd t +\epsilon\Gamma \rd \vB_t,
\label{3b}
\end{equation}
with initial distribution for $\vX^{\dag}(t_n)$ identical to that of $\vX(t_n)$.
\end{subequations}

Recognizing the underlying circulating, conservative dynamics in Eqs. (\ref{3a}) and (\ref{3b})
allows us to connect a Hamiltonian structure with linear stochastic processes, and consequently develop a Helmholtz theorem, which historically has served as the fundamental mathematical link between classical Newtonian mechanics and thermodynamics.  For high dimensional stochastic processes, variables in the Helmholtz theorem provide the systems' underlying dynamics with a macroscopic picture.
An ideal gas-like relation between a set of new, macroscopic
variables emerges, confirming the simplicity of the OUP.
A work-free energy equality in terms of
the macroscopic thermodynamic variables, which are
fluctuating with the underlying dynamics, captures
the nature of the fluctuation in the underlying stochastic processes.
We emphasize that even though the mathematical derivations
are essentially the same, the physical meaning of the work
relation is closer to the classical thermodynamics.

	The paper is structured as follows.  In Sec. \ref{sec:1}, we
first provide the necessary preliminaries on the OUP.
Sec. \ref{sec:1.1} introduces the conservative dynamics as
a part of the stationary behavior of the OUP.
Sec. \ref{sec:1.2} then discusses a long neglected issue of
zero energy reference.  Secs. \ref{sec:2.1} and
\ref{sec:2.2} introduces the stationary free energy
function and the dynamic free energy functional.
Sec. \ref{sec:2.3} studies the novel object of
equation of state.  It is shown that the OUP has a
simply, universal ideal thermodynamic behavior.
In Sec. \ref{sec:3}, we turn to the circulating dynamics
and its relation to classical mechanics as well as
stochastic dynamics.
Sec. \ref{conserve-flow} focuses on the simplicity of the
circulating dynamics as being totally integrable.
Sec. \ref{micro-ensemble} contains a proof
that the stationary probability density of OUP, conditioned
on an invariant torus of the underlying conservative dynamics,
analogous to a microcanonical ensemble, is an invariant measure of the latter.  If the dynamics on an invariant torus is ergodic, then the
conditional probability is the only, natural invariant
measure on the torus.
Work equalities and fluctuation theorems are discussed in
Sec. \ref{sec:4}.  Using a macroscopic presentation of
the Jarzynski equality, its relation to Helmholtz theorem is
revealed in Sec. \ref{macro-jarzy}.  The paper concludes
with discussions in Sec. \ref{sec:discussion}.

\section{Preliminaries}
\label{sec:1}

\subsection{Stationary Gaussian density and underlying
conservative dynamics}
\label{sec:1.1}

The OUP in Eq. (\ref{3a}) satisfies the important
{\em fluctuation-dissipation relation}:
$2 D\Xi^{-1}=\epsilon^2\big(\Gamma\Gamma^T\big)\times$
covariance matrix of the stationary OUP.  In fact, it has a
stationary Gaussian distribution $Z^{-1}(\alpha) e^{-\varphi(\vx;\alpha)/\epsilon^2}$
in which $Z(\alpha)$ is a normalization factor and
$\varphi(\vx;\alpha)=\vx^T\Xi^{-1}(\alpha)\vx$.  In addition, there is
an underlying circulating dynamics
\begin{equation}
           \frac{\rd\vx}{\rd t} = -\Big(M(\alpha)-D\Xi^{-1}(\alpha)\Big)\vx,
\label{conservedy}
\end{equation}
where the scalar $\varphi(\vx;\alpha)$ is
conserved  \cite{kwon_ao_thouless_pnas}:
\begin{equation}
   \frac{\rd}{\rd t}\varphi\big( \vx(t);\alpha \big)
       = -2\vx^T\Xi^{-1}\Big(M-D\Xi^{-1}\Big)\vx
       = -\vx^T\Big(
           \Xi^{-1}M-M^T\Xi^{-1}\Big)\vx = 0.
\end{equation}
In fact, this conservative dynamics can be
expressed as \cite{qian_pla_2014}:
\begin{eqnarray}
   \frac{\rd\vx}{\rd t} = -\frac{1}{2}\Big(M\Xi-D\Big)
                  \nabla_{\vx}\varphi(\vx;\alpha) ,
\label{eq-9}
\end{eqnarray}
where
\begin{eqnarray}
   \varphi(\vx;\alpha) = \vx^T\Xi^{-1}(\alpha)\vx;
\label{eq-E}
\end{eqnarray}
and $\big(M(\alpha)\Xi-D\big)$ is skew-symmetric.

	It is of paramount importance to recall that for
a Markov process without detailed balance, its
stationary dynamics is quantified by two mathematical
objects: a stationary probability density and a stationary
circulation \cite{jqq,wangjin} characterized as a
divergence-free, conservative vector field.
In general, the latter accounts for the complexity arising from the system's dynamics \cite{ChaosYian}: how many integrals of motion does it have;
whether the conservative dynamics is ergodic on an invariant set; etc. Many
of the characteristics persist in the stationary stochastic process, and can be
used to classify long time, complex behaviors in high dimensional systems.
On the other hand, the dissipative (transient) dynamics plus noise drive the system towards the stationary distribution while characterizing
``energy'' fluctuations.

For the OUP in Eq. (\ref{3a}), the conservative dynamics will be shown to be totally integrable.
That is, symmetries would be implied through $\lfloor n/2 \rfloor$ first integrals of motions, which are the natural generalizations of the time-reversal symmetries.
The remaining part, $\rd\vX(t) =$ $-\frac{1}{2}D\nabla_{\vx}\varphi(\vx;\alpha)\rd t +\epsilon\Gamma \rd \vB_t$,
has a stationary dynamics that is
detailed balanced.

It is worth noting that any
$\widetilde{\varphi}(\vx;\alpha)=\varphi(\vx;\alpha)+C(\alpha)$
is also a valid substitution for the $\varphi(\vx;\alpha)$ in
Eq. (\ref{eq-E}). As far as the stochastic dynamical system Eq. (\ref{3a})
is concerned, there is no unique $\widetilde{\varphi}(\vx;\alpha)$ as
a function of both dynamic variable $\vx$ and parameter $\alpha$.

\subsection{Zero energy reference: A hidden assumption in classical physics}
\label{sec:1.2}

	The central object that connects classical Newtonian mechanics
with equilibrium thermodynamics is the entropy function $S(E,V,N)$,
with $V$ and $N$ being the volume and the number of particles
of a classical mechanical system in a container, and $E$
its total mechanical energy which is conserved according to
Newton's Second Law of motion.  In Hamilton's formulation
 Eq. (\ref{1}), $E$ is simply the initial value of the Hamiltonian function
$H\big(\{x_i\},\{y_i\}\big)$ in which $x_i$ and $y_i$ are the position
and momentum of $i$th particle,  respectively, $1\le i\le N$.

	We recognize that in the classical theory of
mechanical motions, replacing $H$ with $\widetilde{H}\big(\{x_i\},\{y_i\}\big)$
$=H\big(\{x_i\},\{y_i\}\big) + C$, where $C$ is a constant, has
absolutely no consequence to the mathematical theory.  Therefore,
with parameters contained in the Hamiltonian function,
such as $V$ and $N$, $H\big(\{x_i\},\{y_i\};V,N\big)$
and $H\big(\{x_i\},\{y_i\};V,N\big)+C(V,N)$ are equivalent.  In other words,
classical mechanics only uniquely determines a Hamiltonian
function up to an arbitrary function of all the non-dynamic
parameters.

		However, an additive function $C(V,N)$ would cause non-uniqueness in the thermodynamic forces in the relation:
\begin{equation}
     \rd S(E,V,N) = \left(\frac{\partial S}{\partial E}\right)_{V,N}
                   \Big[ \rd E  + p\rd V - \mu \rd N
                        \Big],
\label{chem-pot}
\end{equation}
in which
\begin{equation}
     p = \frac{\displaystyle \left(\frac{\partial S}{\partial V}
                   \right)_{E,N}}{\displaystyle \left(\frac{\partial S}{\partial E}
                   \right)_{V,N}  } = - \left(\frac{\partial E}{\partial V}
                   \right)_{S,N},  \   \
      \mu = \left(\frac{\partial E}{\partial N}
                   \right)_{S,V}.
\end{equation}
Corresponding to $\widetilde{H}=H+C(V,N)$ one has, for $E$ as the initial values of $H\big(\{x_i\},\{y_i\};V,N\big)$, and $\widetilde{E}$ as the initial values of $\widetilde{H}\big(\{x_i\},\{y_i\};V,N\big)$:
\begin{equation}
     \widetilde{p} = - \left(\frac{\partial\widetilde{E}}{\partial V}
                   \right)_{S,N} = p
                    -\left(\frac{\partial C}{\partial V}\right)_N,  \   \
      \widetilde{\mu} = \mu + \left(\frac{\partial C}{\partial N}\right)_V.
\end{equation}
Since pressure $p$ has a mechanical interpretation,
one can, by physical principle, uniquely determine the form of
$p$ as a function of $V$.  The situation for $\mu$ is much
less clear:  Since there is not an independent mechanical
interpretation of the chemical potential other than
the thermodynamic one given in Eq. (\ref{chem-pot}), the
non-uniqueness is inherent in the mathematical,
as well as the physico-chemical theory.   The problem
has the same origin as {\em Gibbs' paradox}
\cite{noyes,ge_qian_gp}.

	In classical chemical thermodynamics, the
Hamiltonian function as a function of varying
number of particles $N$, $H(x_1,\cdots,x_N,y_1,\cdots,y_N)=$
$\frac{1}{2}\sum_i^N m_iy_i^2+V(x_1,\cdots,x_N)$, is
uniquely determined via a Kirkwood
charging process \cite{k-charging}:
\begin{equation}
     V\big(x_1,\cdots,x_N,x_{N+1}\big) =
     V\big(x_1,\cdots,x_N\big)+ \sum_{j=1}^N U_{j,N+1}\big(x_j,x_{N+1}\big),
\end{equation}
where
\begin{equation}
     \lim_{|x-y|\rightarrow\infty} U(x,y) = 0.
\end{equation}
With this convention, the Hamiltonian for a molecular system is
uniquely determined in chemical thermodynamics, which yields
a consistent chemical potential $\mu$.  How to generalize
this chemical approach to Hamiltonian dynamics
Eq. (\ref{1}) with no clear separation between kinetic and
potential parts, however, is unclear.

	The problem of uniqueness of Hamiltonian
function $\widetilde{H}$ is intimately related to the uniqueness of
$\widetilde{\varphi}(\vx;\alpha)$ in Sec. \ref{sec:1.1}.
As we shall show in the rest of this paper, the
zero energy reference has deep implications to
the theory of stochastic
thermodynamics.  The resolution to the problem will be
discussed in Sec. \ref{sec:resol}.

\section{Free energy functions and functional}
\label{sec:2}

As the notion of entropy, the definition of free energy is widely varied
in the literature.\footnote{It has become
increasingly clear that the Boltzmann's entropy for a
Hamiltonian dynamics is not unique: There are different
geometric characterizations of the level sets of the Hamiltonian
that can be acceptable choices.  Neither is Shannon's entropy in
stochastic dynamics unique: other convex functions such as
Tsallis' entropy can also be found in the literature.}  The most general features of free energy, perhaps,
are:  it is the difference between ``internal energy'' and
entropy; it is the entropy under a ``natural invariant measure''.
In this section, we shall present two different types
of free energies associated with the OU dynamics in Eq. (\ref{3a}):

($i$) Thermodynamic free energy of a stationary dynamics, as a
function of mean internal energy $E$ and parameter $\alpha$:
$A(E,\alpha)$.  We identify a ``thermodynamic state'' as a
state of sustained motion, either for a deterministic conservative
dynamics Eq. (\ref{conservedy}), or for a stochastic stationary process
defined by Eq. (\ref{3a}).

($ii$) Dynamic free energy functional, $\Psi[f(\vx,t)]$,
for an instantaneous probability distribution $f(\vx,t)$.

\subsection{Thermodynamic free energy functions $A(E,\alpha)$}
\label{sec:2.1}

	With a particular given $\varphi(\vx;\alpha)$, we now introduce
two different free energy functions.
The first one is defined following the microcanonical ensemble approach;
definition of the second one follows Gibb's canonical ensemble approach.
While the second one is frequently being used in the work-free energy relation
(discussed in Sec. 5), the two definitions agree perfectly in the large dimension limit.

The first thermodynamic free energy function, $A_1(E,\alpha)$, associated
with the conservative deterministic motion of Eq. (\ref{conservedy})
on the surface of $\varphi(\vx;\alpha)=E$, is obtained
following the microcanonical ensemble approach through Boltzmann's
entropy function.
Letting $\sigma_B(E,\alpha)$ correspond to the entropy $S$
and $\Theta^{-1}(E,\alpha)$ correspond
to $\left(\frac{\partial S}{\partial E}\right)$
in Eq. (\ref{chem-pot}), we can define:
\begin{subequations}
\label{b-approach}
\begin{eqnarray}
	\sigma_B(E,\alpha) &=& \ln
            \left( \int_{\varphi(\vx;\alpha)\le E}
           \rd\vx \right)
\nonumber\\
	&=& \frac{n}{2}\ln E + \frac{1}{2}\ln \det\Xi(\alpha) + \ln V_n,
\\
	\Theta^{-1}(E,\alpha) &=& \left(\frac{\partial\sigma_B}{\partial E}
                \right)_{\alpha} = \frac{n}{2E},
\\
	A_1(E,\alpha) &=&  E - \Theta\sigma_B
\nonumber\\
	&=&  \frac{2E}{n}
      \left\{-\frac{n}{2}\ln E - \frac{1}{2}\ln \det\Xi(\alpha) - \frac{n}{2}
       \ln (\pi) +\ln\Gamma\left(\frac{n}{2}+1\right) + \frac{n}{2} \right\},  \label{A1} \nonumber\\
\end{eqnarray}
\end{subequations}
where $V_n=\pi^{n/2}\left(\Gamma\left(\frac{n}{2}+1\right)\right)^{-1}$
is the volume of an $n$-dimensional Euclidean ball with radius 1.
$\Gamma(\cdot)$ is gamma function. $n$ is the dimension of the
OUP in Eq. (\ref{3a}).

The second one, $A_2(E,\alpha)$, follows Gibbs' canonical ensemble
approach via the ``partition function'' $Z(\alpha)$:
\begin{subequations}
\label{g-approach}
\begin{eqnarray}
	Z(\alpha) &=& \int_{\mathbb{R}^n} e^{-\varphi(\vx;\alpha)/\epsilon^2}
           \rd\vx   \  =\
           \Big( \big(\pi\epsilon^2\big)^n
                              \det\Xi(\alpha) \Big)^{\frac{1}{2}},
\\
     A_2(\overline{E},\alpha)
        &=&  -\epsilon^2\ln Z(\alpha)
\nonumber\\
         &=&  \frac{2\overline{E}}{n}
            \left\{-\frac{n}{2}\ln \overline{E} -\frac{1}{2}\ln\det\Xi(\alpha)
           -\frac{n}{2}\ln (\pi)
           +\frac{n}{2}\ln\left(\frac{n}{2}\right) \right\},
\end{eqnarray}
in which mean internal energy
\begin{equation}
   \overline{E}= \frac{1}{Z(\alpha)}
             \int_{\mathbb{R}^n} \varphi(\vx;\alpha)
              e^{-\varphi(\vx;\alpha)/\epsilon^2} \rd\vx
            =\frac{1}{2}n\epsilon^2.
\end{equation}
\end{subequations}

	The two free energy functions $A_1$ in Eq. (\ref{b-approach}c) and $A_2$ in Eq. (\ref{g-approach}b) are
different only by a function of $n$ inside the $\{\cdots\}$.
For large $n$, $\ln\Gamma(\frac{n}{2}+1)\approx \frac{n}{2}\ln\left(
\frac{n}{2}\right)-\frac{n}{2}$.  Therefore, $A_1$ and $A_2$
agree perfectly in the limit of $n\rightarrow\infty$.

\subsection{Dynamic free energy functional $\Psi[f_{\alpha}(\vx,t)]$}
\label{sec:2.2}

	The thermodynamic free energy $A_2(E,\alpha)$ in
Sec. \ref{sec:2.1} sets a universal energy reference point
for the entire family of stochastic dynamics in Eq. (\ref{3a}) with
different $\alpha$.  For a given $\alpha$, the time-dependent
probability density function  $f_{\alpha}(\vx,t)$ follows the partial differential
equation
\begin{equation}
   \frac{\partial f_{\alpha}(\vx,t)}{\partial t} = \nabla_{\vx}\cdot\Big(\epsilon^2 D\nabla_{\vx} f_{\alpha}(\vx,t)
              +M(\alpha)\vx f_{\alpha}(\vx,t) \Big).
\label{fpe}
\end{equation}
The $f_{\alpha}(\vx,t)$ represents an instantaneous ``state'' of the
probabilistic system, which has a free energy functional
\begin{eqnarray}
    \Psi\big[f_{\alpha}(\vx,t)\big]  &=&  \int_{\mathbb{R}^n}
        \varphi(\vx;\alpha) f_{\alpha}(\vx,t) \ \rd\vx
            -\left(-\epsilon^2\int_{\mathbb{R}^n} f_{\alpha}(\vx,t)
       \ln f_{\alpha}(\vx,t) \ \rd\vx \right)
\nonumber\\
      &=&  \epsilon^2\int_{\mathbb{R}^n} f_{\alpha}(\vx,t)
       \ln\left( \frac{ f_{\alpha}(\vx,t)}{
             Z^{-1}(\alpha)e^{-\varphi(\vx;\alpha)/\epsilon^2}
               }\right)\rd\vx + A_2(\alpha).
\end{eqnarray}
This is a dynamic generalization of the free energy functions
in Sec. \ref{sec:2.1}.  It has two important properties.  First,
\begin{equation}
    \lim_{t\rightarrow\infty}  \Psi\big[f_{\alpha}(\vx,t)\big] = A_2(\alpha).
\end{equation}
Second \cite{qian_pla_2014,qian_epj_st},
\begin{equation}
   \frac{\rd}{\rd t}  \Psi\big[f_{\alpha}(\vx,t)\big] \le 0,
\label{eq-14}
\end{equation}
in which the equality holds if and only if $f_{\alpha}(\vx,t)$ reaches its stationary
distribution $Z^{-1}(\alpha)e^{-\varphi(\vx;\alpha)/\epsilon^2}$.
The negated rate of change in the dynamic free energy functional, $-\rd\Psi/\rd t$, is widely recognized as non-adiabatic
entropy production rate.

	The entropy production rate also has a finite time, stochastic
counterpart in terms of the logarithm of the likelihood ratio:
\begin{equation}
	  -\frac{\rd}{\rd t}\Psi\big[f_{\alpha}(\vx,t)\big] = \lim_{s\rightarrow t}
       E^{\mathbb{P}}\left[ \frac{1}{|s-t|}
           \ln\left( \frac{f\big( \vX(\tau)|t\le\tau\le s \big) }{
           f_{\vX^{\dag}}\big(\hat{\vX}(\tau)|t\le \tau\le s \big) }   \right) \right],
\end{equation}
where $\hat{\vX}(\tau)=\vX(s-\tau+t)$,
and the expectation $E^{\mathbb{P}}\big[\cdots\big]$ is carried out over
the diffusion process defined by Eq. (\ref{3a}) and the corresponding Eq. (\ref{fpe}):
\begin{equation}
      f\big( \vX(\tau)|t\le\tau\le s \big)
      \propto \exp\left[-2\int_t^s  \Big(\dot{\vX}(\tau)+M\vX(\tau)\Big)^T D
                  \Big(\dot{\vX}(\tau)+M\vX(\tau)\Big) \rd\tau \right].
\end{equation}

\subsection{Universal equation of state of OU process}
\label{sec:2.3}

With the introduction of the internal energy $E$ and the parameter $\alpha$,
the thermodynamic relation Eq. (\ref{chem-pot}) - known as the Helmholtz theorem - for the OUP model can be expressed by
$\sigma_B$, $\alpha$ and their conjugate variables.
We notice that $\alpha$ enters Eq. (\ref{b-approach}) only through
$\det\Xi(\alpha)$.
If one measures $\alpha$ through
$\widetilde{\alpha}=\det\Xi(\alpha)$,
then the Helmholtz theorem writes:
\begin{align}
\rd E &= \Theta(E,\widetilde\alpha) \rd \sigma_B
- F_{\widetilde\alpha}(E,\widetilde\alpha) \rd \widetilde\alpha \nonumber\\
&= \left(\dfrac{\partial \sigma_B}{\partial E}\right)^{-1} \rd \sigma
- \left(\dfrac{\partial \sigma_B}{\partial \widetilde{\alpha}}\right)\left(\dfrac{\partial \sigma_B}{\partial E}\right)^{-1} \rd \widetilde{\alpha}.
\end{align}
The two conjugate variables,
$\Theta$ and $F_{\widetilde{\alpha}}$, correspond to the macroscopic quantities in classical thermodynamics as temperature and force.\footnote{The force here should
be understood as Onsager's thermodynamic force: corresponding
to a spatial displacement is a mechanical force; to a change in number of
particles is Gibbs' chemical potential; to a variation in a parameter through
a Maxwell demon then is an informatic force \cite{sano,jarzynski_md}. }

	Following either Boltzmann's microcanonical or
Gibbs' canonical approach, Sec. \ref{sec:2.1} revealed that
$E=\frac{1}{2}n\Theta$ in which $\theta=\frac{1}{2n}\Theta$ could be
interpreted as an ``absolute temperature''.
Since the absolute temperature $\theta$ is a fluctuating quantity with respect to $E$ and $\widetilde{\alpha}$, it may, in general, not bear a simple relationship with the noise strength $\epsilon^2$. But here in OUP, by comparing the microcanonical approach with the canonical one, we note that the mean absolute temperature $\bar\theta=\frac{\epsilon^2}{2n}$.

The thermodynamic conjugate variable of $\widetilde\alpha$, the $\widetilde\alpha$-force:
\begin{equation}
    F_{\widetilde{\alpha}} = \Theta
   \left(\frac{\partial\sigma_B(E,\widetilde{\alpha})}{\partial\widetilde{\alpha}}\right)_E
        = \frac{n\theta}{\widetilde{\alpha}}.
\end{equation}
A mathematical relation between
$\widetilde{\alpha}$, $F_{\widetilde{\alpha}}$,
and $\theta$ is called an  {\em equation
of state} in classical thermodynamics.

	The ``internal energy'' $E$ being a sole function of temperature
$\theta$, and the product of thermodynamic conjugate
variables,
$\widetilde{\alpha}F_{\widetilde{\alpha}}$,
equaling to $n\theta$, are hallmarks of thermodynamic
behavior of ideal gas and ideal solution.  We thus conclude
that the OUP has a universal ideal thermodynamic behavior.

\section{Circulating conservative flow and its invariant measures}
\label{sec:3}

After discussing the energy function and stationary probability, we now focus
on the dynamic complexity of the system and study the circulating, conservative dynamics.
The universal ideal thermodynamic behavior reveals one aspect of the
simplicity in OUP; another is reflected in the divergence-free motions.
For the linear conservative dynamics, Eq. (\ref{conservedy}), its structure is
known to be simple: the vector field is integrable.

	The conservative dynamics in Eq. (\ref{conservedy}) can be proved to be
purely cyclic (e.g., periodic, or quasi-periodic on an invariant torus).
Because the skew-symmetric matrix $(M-D\Xi^{-1})$ has only pairs of imaginary
eigenvalues $\{\lambda_{\ell}|1\le\ell\le n\}$.
We can also find real Jordan form of $(M-D\Xi^{-1})$:
$Q J Q^{-1}$, where $J$ is block diagonal, with $2\times 2$ skew-symmetric blocks:
\[
 {\rm Im}\left[\lambda_{(2 i - 1)}\right]
\left(
\begin{array}{ll}
         0 & 1 \\
         -1 & 0
\end{array}
\right)
\]
being the $i$th block on the diagonal.
Natural coordinates for the conservative flow Eq. (\ref{conservedy}) is therefore: $\mathbf{y}=Q^{-1}\mathbf{x}$.

\subsection{The conservative flow and general time reversal symmetries}

\label{conserve-flow}

Poisson bracket $\{\cdot,\cdot\}$ can be defined for the linear conservative system as:
$\{\varphi(\mathbf{x}),\psi(\mathbf{x})\}
= \nabla\varphi(\mathbf{x})^T (M-D\Xi^{-1})\nabla\psi(\mathbf{x})$.
Then the conservative flow expressed in terms of its Hamiltonian
function $\varphi(\vx)$ is:
\begin{equation}
\dot{x_i} = \big\{x_i,\dfrac12\varphi(\vx)\big\}.
\end{equation}
First integrals $I_i$ of the conservative flow are:
\begin{eqnarray}
I_i
= y_{2i-1}^2 + y_{2i}^2
= \mathbf{x}^T Q^{-T} I_{(2i-1)\sim(2i)} Q^{-1} \mathbf{x}, \  \
        1\leq i \leq \left\lfloor\dfrac{n}{2}\right\rfloor.
\end{eqnarray}
Here, $I_{(2i-1)\sim(2i)}$ denotes the diagonal matrix with $1$ on $(2i-1)$-th to $(2i)$-th diagonal entries, and zero everywhere else.

The conservative flow is totally integrable, and can be written in canonical action-angle variables.
Angular coordinates $\theta_i$ accompanying $I_i$ can be found as:
\begin{eqnarray}
\theta_i =
{\rm Im}\big[\lambda_{2 i - 1}\big]^{-1}
\cdot \arctan\left(\dfrac{y_{2i-1}}{y_{2i}}\right), \  \
  1\leq i \leq \left\lfloor\dfrac{n}{2}\right\rfloor.
\end{eqnarray}
Hence, in the canonical action-angle variables,
$\varphi=\sum\limits_{i=1}^{\lfloor n/2 \rfloor} I_i$,
\begin{eqnarray}
\left\{
\begin{array}{l}
         \theta_i' = \dfrac{\partial\varphi}{\partial I_i} = 1 \\[10pt]
         I_i' = - \dfrac{\partial\varphi}{\partial \theta_i} = 0.
        \end{array}
\right.
\end{eqnarray}
There are $\left\lfloor\frac{n}{2}\right\rfloor$ first integrals, but for the given Poisson bracket, one combination of them is unique, which is the Hamiltonian $\varphi$ that connects to the stationary distribution and generates the conservative flow.

In the action-angle variables, it is observable that the system bears the following symmetries:
$(t, \theta_i, I_i)\longrightarrow \big(-t, -\theta_i, (-1)^{k_i} I_i \big)$, where $\{k_i\}$ is a sequence of $0$ and $1$.
Taking $\{k_i\}$ as a sequence of zeros, we recover the time-reversal invariance in Eq. (\ref{1}).
Hence, for general $\{ k_i \}$, those symmetries are the natural generalizations of the time-reversal symmetry, as displayed in classical Hamiltonian systems.

\subsection{Conditional probability measure as invariant measure of the conservative flow}
\label{micro-ensemble}

	The OUP yields an equilibrium probability density function for
$\vX$: $f_{\vX}^{eq}(\vx;\alpha)=$
$Z^{-1}(\alpha)e^{-\varphi(\vx;\alpha)/\epsilon^2}$.
In this section, we calculate the {\em conditional probability density}
for $\vX$ restricted on an equal energy surface $\mathfrak{D}_{\varphi=E}=\big\{\vx|\vx\in\mathbb{R}^n,\varphi(\vx;\alpha)=E\big\}$
and prove it to be an invariant measure of the conservative dynamics, Eq. (\ref{conservedy}), restricted
on $\mathfrak{D}_{\varphi=E}$.
Therefore, in the absence of fluctuation and dissipation, our definition of ``equilibrium free energy" (Eq. (\ref{A1})) in stochastic thermodynamics retreats to the Boltzmann's microcanonical ensemble approach in classical mechanics.

One can obtain a conditional probability density
for $\vX$ restricted on an equal energy surface
$\mathfrak{D}_{\varphi=E}=\big\{\vx|\vx\in\mathbb{R}^n,\varphi(\vx;\alpha)=E\big\}$ as:
\begin{equation}
         Z(\alpha) = \int_{\mathbb{R}^n}e^{-\varphi(\vx;\alpha)/\epsilon^2}
               \rd \vx
        = \int_{\varphi_{min}(\alpha)}^{\infty} \exp\left( -\frac{E}{\epsilon^2}
               + S(E,\alpha) \right) \rd E,
               \label{free-energy}
\end{equation}
in which \cite{khinchin}
\begin{equation}
	S(E,\alpha) = \ln\left(\frac{\partial}{\partial E}
        \int_{\varphi(\vx;\alpha)\le E}
            \rd\vx
              \right)_{\alpha} =  \ln\left(
        \oint_{\varphi(\vx;\alpha)= E}
               \frac{\rd \Sigma^{n-1}}{\|\nabla_{\vx}\varphi(\vx;\alpha)\|}
              \right).
\end{equation}
The conditional probability density at $\vx\in\mathfrak{D}_{\varphi=E}$
is:
\begin{equation}
		 \dfrac{ e^{-S(E,\alpha)} }{ \|\nabla_{\vx}\varphi(\vx;\alpha)\|}
           =  \frac{1}{\frac{n}{2}V_nE^{\frac{n}{2}-1}\big(\det\Xi
                    \big)^{-\frac{1}{2}} \|\Xi^{-1}\vx\|}.
\end{equation}
Note this conditional probability is one of the invariant
measures of the conservative dynamics Eq. (\ref{conservedy}) restricted
in $\mathfrak{D}_{\varphi=E}$.

To prove this fact, define the dynamics of the conservative part as: ${\bf S}_t$, mapping a measurable set $A\rightarrow {\bf S}_t(A)$.
Then measure of a set $A \subseteq \mathfrak{D}_{\varphi=E}$ under
$\rd \mu= e^{-S(E,\alpha)} \|\nabla_{\vx}\varphi(\vx;\alpha)\|^{-1} \rd \Sigma^{n-1}$ is:
\begin{align}
\int_A \dfrac{e^{-S(E,\alpha)}}{\|\nabla_{\vx}\varphi(\vx;\alpha)\|} \rd \Sigma^{n-1}
&= e^{-S(E,\alpha)} \int_A \delta(E-\varphi(\vx)) \rd \vx \nonumber\\
&= e^{-S(E,\alpha)} \int_{{\bf S}_t^{-1}(A)} \delta(E-\varphi(\vx)) \rd \vx \nonumber\\
&= \int_{S_t^{-1}(A)} \dfrac{e^{-S(E,\alpha)}}{\|\nabla_{\vx}\varphi(\vx;\alpha)\|} \rd \Sigma^{n-1},
\end{align}
since ${\bf S}_t^{-1}(A)\subseteq \mathfrak{D}_{\varphi=E}$ and ${\bf S}_t$ is volume preserving.
In general, if the dynamics is ergodic on the entire $\mathfrak{D}_{\varphi=E}$,
then its invariant measure $\mu$ is the {\em physical measure}: $\mu$-average
equals time average along a trajectory;
if there are other first integrals for the conservative dynamics,
then $\mu$ can be projected further to lower dimensional invariant sets.

\subsection{Resolutions to the energy reference problem}
\label{sec:resol}

Up to now, there are clearly several possibilities to uniquely
determine the free additive function $C(V,N)$ in the Hamiltonian $H\big(\{x_i\},\{y_i\};V,N\big)$ discussed in Introduction.

(1)  $C(V,N)$ is chosen such that the global
minimum of $H=0$ for each and every $V$ and $N$.
This is widely used, implicitly, in application
practices, as in our Eq. (\ref{eq-E}).

(2) $C(V,N)$ is chosen according to the ``equilibrium
free energy'':
\begin{equation}
         - \epsilon^2 \ln \int e^{-E/\epsilon^2+S(E,V,N)}\rd E
                  = 0.
 \end{equation}
Note that this is precisely the ``energy function''
in Hatano and Sasa \cite{sasa}.

(3)  Extra information concerning the fluctuations in
$V$, such as in an isobaric ensemble, and
fluctuations in $N$ in grand ensemble, provides an
empirically determined basis for the free energy scale.

In terms of the theory of probability, choice (2) uniquely
determines the energy reference point according to a
conditional probability, and in choice (3) it is uniquely
determined according to a marginal distribution.
How to normalize a probability, which has always been
considered non-consequential in statistical physics, seems to
be a fundamental problem in the physics of complex systems.
%
%

\section{Work equalities and fluctuation theorems}
\label{sec:4}

The previous discussions suggest that while a great deal of
complexity of a detailed, mesoscopic stationary dynamics is captured
by the circulating conservative dynamics, OUP also has a
macroscopic state of motion that is defined by the internal energy $E$,
or equivalently the level sets of $\varphi(\vx)$,$\mathfrak{D}_{\varphi=E} \subset \mathbb{R}^n$.
Thus, from the macroscopic point of view, a stochastic system could be studied through the one-dimensional (1-D) time sequence of fluctuating internal energy $E$, as a function of $t$, or the
change in $E$ due to changes in the parameter $\alpha$.

The celebrated Jarzynski equality connects the mesoscopic fluctuating
force with the change in free energy.  We present this result through
a projection from $n$-D phase space to $1$-D function $E(t)$ that facilitates experimental verification of the work-free energy relation.
This approach reveals a close connection between the Jarzynski equality and the Helmholtz theorem.
We start with stating the Jarzynski and Crooks' equalities,
with the mathematical proofs collected in the Appendix \ref{app-A}
for readers' convenience. We then demonstrate the novel formulation of the Jarzynski equality
in the projected space.

	We have shown that $A_2(\alpha)$ is uniquely determined
only up to a particular $\varphi(\vx;\alpha)$.  As shown below, the existence of
an $A_2(\alpha)$ has a paramount importance in the theories
of work equalities, in which the notion of a common
energy for a family of stochastic dynamical systems with
different $\alpha$ has to be given {\it a priori} \cite{sekimoto-book}.

\subsection{The Jarzynski equality}

	The macroscopic $\alpha$-force in the Helmholtz
theorem, as a function of $E$ and $\alpha$, is defined through Boltzmann's entropy $\sigma_B$.  The Jarzynski equality, on the
other hand, concerns with a mesoscopic $\alpha$-force,
\begin{equation}
   F_{\alpha}(\vx;\alpha) = -\left(\frac{\partial\varphi(\vx;\alpha)}{\partial\alpha}\right)_{\vx},
\end{equation}
and the statistical behavior of its corresponding {\em stochastic work}
\begin{equation}
   W[\vX(\tau),\alpha(\tau)] =
    \int_0^t  F_{\alpha}\big(\vX(\tau);\alpha(\tau)\big)
          \left(\frac{\rd\alpha(\tau)}{\rd\tau}\right) \rd \tau.
\end{equation}
The Jarzynski equality dictates that if the initial
distribution of $\vX(\tau)$ follows the equilibrium distribution, then \cite{jarzynski}
\begin{align}
\left\langle e^{ -\frac{1}{\epsilon^2}
        W[\vX(\tau),\alpha(\tau)] } \right\rangle_{\big[\vX(\tau),\alpha(\tau)\big]}
= e^{-\frac{1}{\epsilon^2}\Delta A_2(\alpha) },
\end{align}
where the average of a functional over the ensemble of paths is defined as:
\begin{align}
\Big\langle G[\vX(\tau),\alpha(\tau)]\Big\rangle_{\big[\vX(\tau),\alpha(\tau)\big]}
= \int G[\vX(\tau),\alpha(\tau)] \mathcal{P}[\vX(\tau),\alpha(\tau)]
                 \mathcal{D}[\vX(\tau)],
\label{avg}
\end{align}
in which $\mathcal{P}[\vX(\tau),\alpha(\tau)]\mathcal{D}[\vX(\tau)],$ is
an infinite-dimensional probability distribution
for the entire paths $[\vX(\tau)]$.

It is clear from the proof in Appendix \ref{app-A} that the Jarzynski equality is general for Markov processes with or without detailed balance.  Several recent papers have studied extensively the latter case \cite{tangying,chulan-kwon}.

\subsection{Crooks' approach}
G. E. Crooks' approach, when applied to processes without detailed
balance \cite{sasa}, considers the probability functional of a backward path
$\mathcal{P}[\check{\vX}(t)|\check{\vX}(0);\check{\alpha}(t)]$
over a forward one $\mathcal{P}[\vX(t)|\vX(0);\alpha(t)]$,
where both the initial and final
distribution of $\vX(\tau)$ follows the equilibrium distribution:
\begin{align}
\dfrac{\mathcal{P}[\check{\vX}(\tau),\check{\alpha}(\tau)]}
{\mathcal{P}[{\vX}(\tau),{\alpha}(\tau)]}
= &\exp\left( \frac{ Q[\vX(\tau),\alpha(\tau)]
- Q_{hk}[\vX(\tau),\alpha(\tau)]
- \Delta \varphi + \Delta A_2 }{ \epsilon^2 } \right) \nonumber\\
= &\exp\left( \frac{- W[\vX(\tau),\alpha(\tau)]
- Q_{hk}[\vX(\tau),\alpha(\tau)] + \Delta A_2 }{\epsilon^2}\right) , \label{PathProb}
\end{align}
where
$\check{\vX}(\tau) = \vX(t-\tau)$, $\check{\alpha}(\tau) = \alpha(t-\tau)$;
and
\begin{align}
&Q[\vX(\tau),\alpha(\tau)]=\int_0^t \dfrac{\partial \varphi}{\partial \vX} \dot{\vX} \ \rd \tau,
\nonumber\\
&Q_{hk}[\vX(\tau),\alpha(\tau)]
={\displaystyle -2\int_0^t \dot{\vX}^T D(\alpha)^{-1}
\big(M(\alpha) - D(\alpha) \Xi^{-1}(\alpha)\big) \vX \ \rd \tau}
\end{align}
are the heat dissipation and the house-keeping heat respectively.

If a process describes a physical system in
equilibrium, which is expected to be ``microscopic reversible'' in \cite{crooks}, then
\begin{align}
\left<\dfrac{\mathcal{P}[\check{\vX}(\tau),\check{\alpha}(\tau)]}
{\mathcal{P}[{\vX}(\tau),{\alpha}(\tau)]}\right>_{\big[{\vX}(\tau),{\alpha}(\tau)\big]}
= \int \mathcal{D} [{\vX}(\tau),{\alpha}(\tau)] \
\mathcal{P}[\check{\vX}(\tau),\check{\alpha}(\tau)]
= 1.
\end{align}
On the other hand, if the system is in detailed balance for each and
every $\alpha$,
$M(\alpha) - D(\alpha) \Xi^{-1}(\alpha)=0$, then
the house-keeping heat $Q_{hk}[\vX(\tau),\alpha(\tau)]\equiv 0$.
Therefore, path-ensemble average of Eq. (\ref{PathProb}) gives:
\begin{eqnarray}
1  &=& e^{\Delta A_2/\epsilon^2}
\left<e^{  -W[{\vX}(\tau),{\alpha}(\tau)]/\epsilon^2
- Q_{hk}[\vX(\tau),\alpha(\tau)]/\epsilon^2 }\right> _{\big[{\vX}(\tau),{\alpha}(\tau)\big]} \nonumber\\
 &=& e^{\Delta A_2/\epsilon^2}
\left<e^{-W[{\vX}(\tau),{\alpha}(\tau)]/\epsilon^2 }\right> _{\big[{\vX}(\tau),{\alpha}(\tau)\big]}.
\label{CrooksEq}
\end{eqnarray}
This is Hatano-Sasa's result \cite{sasa}.
For systems without detailed balance, $Q_{hk}[\vX(\tau),\alpha(\tau)]$ measures the magnitude of the divergence-free vector field, or the extent to which the system is away from detailed balance, even when stationary distribution is attained.
At the same time,
\[
     \left<\dfrac{\mathcal{P}[\check{\vX}(\tau),\check{\alpha}(\tau)]}
{\mathcal{P}[{\vX}(\tau),{\alpha}(\tau)]}\right>_{\big[{\vX}(\tau),{\alpha}(\tau)\big]}
\]
measures how much on average the behavior of backward paths is statistically different from forward ones.

\subsubsection{Crooks' approach through adjoint processes}

	Jarzynski's approach is based on a mesoscopic $\alpha$-force;
while Crooks' approach concerns with the stochastic entropy
production rate which reflects ``heat dissipation''.  Therefore,
for systems with detailed balance,
they are essentially the same result according to the First Law
of thermodynamics.  For systems without detailed balance,
one can again obtained a Jarzynski-like equality
from the probability $P$ of the forward path over the adjoint
probability $P^\dag$ of the backward one,
according to the notion of {\em time reversal} in Eq. (\ref{3b}):
\begin{align}
P^\dag(\vX(\tau)|\vX(\tau+\rd \tau);\alpha(\tau)) = &P(\vX(\tau+\rd \tau)|\vX(\tau);\alpha(\tau))
\nonumber\\
&\times \left( \dfrac{f_{\vX}^{eq}(\vX(\tau);\alpha(\tau))}
{f_{\vX}^{eq}(\vX(\tau+\rd \tau);\alpha(\tau))}\right).
\end{align}
Thus, whether a system is in detailed balance or not, one has
the Hatano-Sasa equality \cite{sasa}:
\begin{align}
1 = \left\langle\dfrac{\mathcal{P}[{\vX}(\tau),{\alpha}(\tau)]}
{\mathcal{P}^\dag[\check{\vX}(\tau),\check{\alpha}(\tau)]}\right\rangle_{\big[\vX(\tau),\alpha(\tau)\big]}
= e^{\Delta A_2/\epsilon^2 }
\left\langle e^{-W[\vX(\tau),\alpha(\tau)]/\epsilon^2 }\right\rangle_{\big[\vX(\tau),\alpha(\tau)\big]}.
\end{align}

\subsection{Macroscopic work equalities}
\label{macro-jarzy}

We are now in the position to study the work-free energy relation from a macroscopic view.  Essentially, we will consider the stochastic, fluctuating $\left(E(t),\widetilde\alpha(t)\right)$ instead of $\left(\vX(t);\alpha(t)\right)$ directly.
In doing so, we are observing the evolution in the probability distribution of $E$ through a projection from $(\vX;\alpha)$ to $(E,\widetilde\alpha)$.
With the projection of the $n$-dimensional phase space to the  one-dimensional time series $E(t)$,
the stationary probability density function $f^{ss}_{\bf E}(E,\widetilde\alpha)$ of $E$ with $\widetilde\alpha$ is also a projection of the original stationary probability density function $f^{ss}_\vX(\vx;\widetilde\alpha)$ in Euclidean space (as discussed in Sec. \ref{micro-ensemble}):
\begin{align}
f^{ss}_{\bf E}(E,\widetilde\alpha)
&= \oint_{\varphi(\vx;\widetilde\alpha) = E}
\dfrac{f^{ss}_{\bf X}(\vx;\widetilde\alpha) \ \rd \Sigma^{n-1}}
{||\nabla_\vx \varphi(\vx;\widetilde\alpha)||} \nonumber\\
&= \dfrac{1}{Z(\widetilde\alpha)} \exp\left( -\dfrac{E}{\epsilon^2} + S(E,\widetilde{\alpha})\right),
\end{align}
in which
\begin{align}
S(E,\widetilde\alpha)
= \ln \left(\oint_{\varphi(\vx;\widetilde\alpha) = E}
\dfrac{\rd \Sigma^{n-1}}
{||\nabla_\vx \varphi(\vx;\widetilde\alpha)||}\right).
\label{S_entropy}
\end{align}
For the process of $\left(E(t),\widetilde\alpha(t)\right)$, the total internal energy is no longer $E$ itself.
But rather, it would include the ``entropic effect", $S(E,\widetilde\alpha)$, caused by the curved space structure,
and become $\mathscr{A}(E,\widetilde\alpha)$, as will be discussed
in more detail in \cite{ma-qian-ye}:
\begin{align}
\mathscr{A}(E,\widetilde\alpha)
= E- \epsilon^2 S(E,\widetilde\alpha).
\end{align}

The Helmholtz theorem for the new $\left(E(t),\widetilde\alpha(t)\right)$ process reads:
\begin{align}
\rd \mathscr{A}
&= \widetilde\Theta(E,\widetilde\alpha) \rd \sigma
- \widetilde{F}_{\widetilde\alpha}(E,\widetilde\alpha) \rd \widetilde\alpha \nonumber\\
&= \left(\left(\dfrac{\partial \sigma}{\partial E}\right)^{-1}
- \epsilon^2 \dfrac{\partial S}{\partial \sigma}\right) \rd \sigma
- \left(\left(\dfrac{\partial \sigma}{\partial \widetilde\alpha}\right)\left(\dfrac{\partial \sigma}{\partial E}\right)^{-1}
+ \epsilon^2 \dfrac{\partial S}{\partial \widetilde\alpha}\right) \rd \widetilde\alpha.
\end{align}
Hence, the total force that does the work in this new coordinate is:
\begin{align}
\widetilde{F}_{\widetilde\alpha}(E,\widetilde\alpha)
= - \dfrac{\partial \mathscr{A}(E,\widetilde\alpha)}{\partial \widetilde\alpha}
= \dfrac1n \cdot \dfrac{E}{\widetilde\alpha}
+ \epsilon^2 \dfrac{\partial S(E,\widetilde\alpha)}{\partial \widetilde\alpha} ,
\end{align}
where $\epsilon^2\big(\partial S(E,\widetilde\alpha)/\partial \widetilde\alpha\big)$ is what chemists called an ``entropic force".

Now we define the work that external environment has done to the system through the controlled change of $\widetilde\alpha(t)$ as:
\begin{align}
W[E(\tau),\widetilde\alpha(\tau)]
= - \int_0^t \widetilde{F}_{\widetilde\alpha}(E,\widetilde\alpha) \dot{\widetilde\alpha} \rd \tau.
\end{align}
Then the work-free energy relation in macroscopic variables is:
\begin{align}
\Big\langle e^{-W[E(\tau),\widetilde\alpha(\tau)]}\Big\rangle_{\big[E(\tau),\widetilde\alpha(\tau)\big]}
= \dfrac{Z\big(\widetilde\alpha(t)\big)}{Z\big(\widetilde\alpha(0)\big)}
= \exp\left(-\frac{\Delta A_2(\alpha)}{\epsilon^2}\right).
\end{align}
Therefore, the averaged minus exponential of work is equal to the minus exponential of free energy difference.
Here, we notice that the free energy stays the same through the change of free variables, as a result of Eq. (\ref{free-energy}):
\begin{equation}
         Z(\alpha) = \int_{\mathbb{R}^n}e^{-\varphi(\vx;\alpha)/\epsilon^2}
               \rd \vx
        = \int_{\varphi_{min}(\alpha)}^{\infty} \exp\left( -\frac{E}{\epsilon^2}
               + S(E,\alpha) \right) \rd E. \nonumber
\end{equation}

\section{Discussion}
\label{sec:discussion}

In the present work, using the OUP as an example, we have
illustrated a possible method of deriving emergent, macroscopic
descriptions of a complex stochastic dynamics from its
mesoscopic {\em law of motion}.  In recent years,
there is a growing awareness of the role of probabilistic
reasoning as {\em the} logic of science \cite{jaynes,dill-rmp}.
In this framework, prior information, data, and probabilistic
deduction are three pillars of a scientific theory.   In fields with
very complex dynamics, statistical inferences focus on the latter
two aspects starting with data.   In physical sciences
that includes chemistry, and cellular biology, the prior plays a
fundamental role as a feasible ``mechanism'' which enters
a scientific model based on ``established knowledge'' | no biochemical
phenomena should violate the physical laws of mechanics and
thermodynamics.   Indeed, many priors have  been rigorously formalized
in terms of mathematical theories.   Unfortunately, most of these
theories are expressed in terms of deterministic mathematics
for very simple individual ``particles''; obtaining a meaningful
probabilistic prior for a realistic, macroscopic-level system requires a
computational task that is neither feasible nor meaningful
\cite{pwanderson,aqtw}.   Nonlinear stochastic dynamical study
is the mathematical deductive process that formulates probabilistic
prior based on a given mechanism.

	Open systems, when represented in terms of Markov processes,
are ubiquitously non-symmetric processes according to Kolmogorov's terminology.
This is one of the lessons we learned from the
open-chemical systems theory.  The non-symmetricity can be
quantified by  {\em entropy production} \cite{jqq}.
For discrete-state Markov processes, symmetric processes
are equivalent to
Kolmogorov's cycle condition \cite{kolmogorov_31}.  Interestingly,
concepts such as cycle condition, detailed balance, dissipation and
irreversible entropy production had all been independently
discovered in chemistry: Wegscheider's relation in 1901
\cite{weg_zpc_01}, detailed balance by G.N. Lewis in
1925 \cite{lewis_pnas_25}, Onsager's dissipation function
in 1931 \cite{onsager_31}, and the formulation of
entropy production in the 1940s \cite{prigogine,tolman_rmp}.

	A non-symmetric Markov process implies circulating
dynamics in phase space.  Such dynamics is not necessarily
dissipative, as exemplified by harmonic oscillators in
classical mechanics.  One of us has recently pointed out
the important distinction between {\em overdamped
thermodynamics} and  {\em underdamped thermodynamics}
\cite{qian_pla_2014}.  The present paper is a study
of OUP in terms of the latter perspective, in which we have
identified the unbalanced circulation as a conservative dynamics,
a hallmark of the generalized underdamped thermodynamics \cite{qian_epj_st}.
In terms of this conservative dynamics, Boltzmann's
entropy function naturally enters stochastic thermodynamics,
and we discover a relation between the Helmholtz theorem
\cite{mayian} and the various work relations.

	In the past, studies on stochastic thermodynamics with
underdamped mechanical motions have always required an
explicit identification of even and odd variables. See
recent \cite{imparato-2012} and \cite{qian_pla_2014}
and the references cited within.  One of us has
introduced a more general stochastic formulation of
``underdamped'' dynamics, with thermodynamics, in
which circulating motion can
be a part of a conservative motion \cite{qian_pla_2014}
without dissipation.  The present work is an in-depth
study of the OUP within this new framework.  It seems
to us that even the term ``nonequilibrium'' in the literature
has two rather different meanings:  From a classical
mechanical standpoint, any system with a stationary
current is ``nonequilibrium'', even though it can be
non-dissipative.  From a statistical mechanics
stand point, on the other hand, ``nonequilibrium'',
``irreversible'', and ``dissipative'' are almost all
synonymous.

\subsection{Absolute information theory and interpretive information
theories}

We now discuss two rather different perspectives on the nature
of information theory, or theories \cite{kolmogorov,jaynes}.

First, in the framework of classical physics in
terms of Newtonian mechanics, Boltzmann's law, and
Gibbs' theory of chemical potential, there is a
universal First Law of Thermodynamics based on the
function $S(E,V,N,\alpha)$ where $S$ is the Boltzmann's
entropy of a conservative dynamical system at total
energy $E$, e.g., Hamiltonian $H\big(\{x_i\},\{y_i\}\big)=E$,
with $V$ and $N=(n_1,n_2,\cdots,n_m)$ being the volume and
numbers of particles in the chemomechanical system, and
$\alpha=(\alpha_1,\alpha_2,\cdots,\alpha_{\nu})$ represents
controllable parameters of the system.  Then one has
\begin{equation}
   \rd E = \left(\frac{\partial S}{\partial E}\right)_{V,N,\alpha}^{-1}
 \rd S  - p\rd V + \mu \rd N - \left(\frac{\partial S}{\partial E}\right)_{V,N,\alpha}^{-1}
          \left(\frac{\partial S}{\partial\alpha}\right)_{E,V,N}\rd\alpha,
\label{grand1stlaw}
\end{equation}
in which $(\partial S/\partial E)^{-1}_{V,N,\alpha}$ is absolute
temperature. $p$ and $\mu$ are pressure and chemical
potential, they are the corresponding thermodynamic
forces for changing volume $V$ and number of particles
$N$, respectively.  It is natural to suggest that if an agent is
able to manipulate a classical system through changing $\alpha$
while holding $S$, $V$, and $N$ constant, then he or she is
providing to, or extracting from, the classical system
{\em non-mechanical, non-chemical work}.  It will be the origin of a Maxwell's demon \cite{jarzynski_md}.

	For an isothermal system, one can introduce Helmholtz's
free energy function $A=E-TS$, then Eq. (\ref{grand1stlaw}) becomes
\begin{equation}
    \rd A = \mu\rd N -S\rd T-p\rd V + F_{\alpha}\rd\alpha.
\label{dA}
\end{equation}
And for an isothermal, isobaric information manipulation process
without chemical reactions, one has Gibbs function $G=E-TS+pV$
and $\rd G = \mu\rd N -S\rd T+V\rd p + F_{\alpha}\rd\alpha
= F_{\alpha}\rd\alpha$. Note that while the first three terms
contain ``extensive'' quantities $N$, $S$, and $V$, the last
term usually does not.   It is nanothermodynamic \cite{hill_nano}.
Note also that for a feedback system that controls $F_{\alpha}$,
one has $\Theta = G-F_{\alpha}\alpha$ and
$\rd\Theta = -\alpha\rd F_{\alpha}$.

Just as $\mu$ is a function of temperature
$T$ in general, so is $F_{\alpha}$: It has an entropic
part \cite{hq-jjh,hq-pre-02}.  This is where the ``information'' in Maxwell's demon
enters thermodynamics. Eqs. (\ref{grand1stlaw}) and (\ref{dA}),
thus, are a grander First Law which now
includes feedback information as a part of the conservation \cite{sano}
with ``informatic energy'' $F_{\alpha}\rd\alpha$, on
a par with heat energy $T\rd S$, mechanical energy $p\rd V$,
and Gibbs' chemical energy $\mu\rd N$.  Eq. (\ref{grand1stlaw})
is the theory of absolute information in connection to
controlling $\alpha$.

	In engineering and biological research on complex
systems, however, the notion of information often has a more
subjective meaning, or meanings, usually hidden in the form of
a statistical prior \cite{Polettini1,Polettini2}.   One of the best examples,
perhaps, is in current cellular biology:  Many key biochemical
processes inside a living cell are said to be ``carrying out
cellular signal transduction''.   Various biochemical activities and
changing molecular concentrations are ``interpreted'' as
``intracellular signals'' that instruct a cell to respond to
its environment.  Here, two very different, but complementary,
mathematical theories are equally valid:  Since nearly all
cellular biochemical reactions can be considered at
constant temperature and volume, one describes the
stochastic biochemical dynamics in terms of Gibbs' theory
based on the $\mu$ in Eq. (\ref{dA}).   On the other hand, the same
stochastic biochemical dynamics described in term of the probability
theory can also be represented as an information processing
machine with communication channels and transmissions of
bits of information, carrying out a myriad of biological
functions such as sensing, proofreading, timing, adaptation,
and amplifications of signal magnitude, detection sensitivity,
and response specificity \cite{hhlm}.  The information flow narratives
provide bioscientists a higher level of abstraction of a physicochemical
reality \cite{qian_roy}.
	
	Such an interpretive information theory, however, will lack the
fundamental character of Eqs. (\ref{grand1stlaw}) and (\ref{dA}).
Still, as a multi-scale, coarse grained theory, some inequalities
can be established \cite{santillan-qian,esposito}.  It is also noted
that changing $\alpha$ can always be mechanistically further
represented in terms of changing geometric quantities such as
volume and particle numbers via chemical reactions: The ultimate
physical bases of information and its manipulation have to be
matters and known forces.

	We believe this dual possibility has a fundamental reason,
rooted in Kolmogorov's rigorous theory of probability:
A probability space is an abstract object associated
with which many different  random variables,
as measurements, are possible.   At this point, it is interesting to read
the preface of \cite{jaynes} written by E. T. Jaynes, who is
considered by many as one of the greatest information theorists
since Shannon: ``From many years of experience with its
applications in hundreds of real problems, our views on the
foundations of probability theory have evolved into something
quite complex, which cannot be described in any such simplistic
terms as `pro-this' or `anti-that'.  For example, our system of
probability could hardly be more different from that of Kolmogorov,
in style, philosophy, and purpose.  What we consider to be fully
half of probability theory as it is needed in current applications |
the principles for assigning probabilities by logical analysis
of incomplete information | is not present at all in the Kolmogorov
system.''

	Then in an amazing candidness, Jaynes goes on:
``Yet, when all is said and done, we find ourselves, to our own
surprise, in agreement with Kolmogorov and in disagreement with
its critics, on nearly all technical issues.''

{\bf Acknowledgements.}  We thank Professors P. Ao, M. Esposito, and
H. Ge for helpful advices, Ying Tang, Lowell Thompson, Yue Wang,
and Felix Ye for many discussions.

\appendix

\section{Derivation of work equalities}
\label{app-A}

\subsection{The Jarzynski equality}

	The basic idea for the derivation is as follows:  We
first represent a path $\vX(t)$ by a discrete version with
$N$ steps and write the path probability in terms of the product
of $N$ transition probabilities given by the $\prod_{i=0}^N (\cdots)$
in Eq. (\ref{eq52}).  Then the mean-exponential of negative work
\[
          \left\langle e^{-\frac{1}{\epsilon^2} W[\vX(\tau),\alpha(\tau)]}\right\rangle_{\big[\vX(\tau),\alpha(\tau)
          \big]}
\]
is   \cite{sasa}:
\begin{align}
&\left\langle e^{-\frac{1}{\epsilon^2} W[\vX_0,\cdots,\vX_N ; \alpha_0,\cdots,\alpha_N]}\right\rangle_{\big[\vX_0,\cdots,\vX_N ; \alpha_0,\cdots,\alpha_N\big]}
\nonumber\\
= &\int\cdots\int \prod_{i=0}^N \rd \vX_i \exp\left(
       - \frac{1}{\epsilon^2} \sum_{i=1}^N \Big(\varphi(\vX_i;\alpha_i) - \varphi(\vX_i;\alpha_{i-1})\Big)
             \right)
\nonumber\\
    &  \times \prod_{i=1}^N P\big(\vX_i|\vX_{i-1};\alpha_{i-1}\big) p\big(\vX_0,t_0;\alpha_0\big),
\label{eq52}
\end{align}
in which the work from state $(\vX_i;\alpha_i)$
to state $(\vX_{i+1};\alpha_{i+1})$ is defined as the difference in
the global $\varphi(\vx;\alpha)$ with a common zero reference.  This is
a consequence of the First Law of Thermodynamics.
Since equilibrium is attained at $t_0$, $p(\vx_0;t_0)=f_{\vX}^{eq}(\vx_0;\alpha_0)$.  With the global $\varphi(\vx;\alpha)=-\epsilon^2\ln f_{\vX}^{eq}(\vx;\alpha)-\epsilon^2\ln Z(\alpha)$, we have:
\begin{align}
&\left\langle e^{-\frac{1}{\epsilon^2} W[\vX_0,\cdots,\vX_N ; \alpha_0,\cdots,\alpha_N]}\right\rangle_{\big[\vX_N,\cdots,\vX_0 ; \alpha_N,\cdots,\alpha_0\big]}
\nonumber\\
= &\int\cdots\int \prod_{i=0}^N \rd \vX_i \ \prod_{i=1}^N
\dfrac{f_{\vX}^{eq}(\vX_i;\alpha_i) Z(\alpha_i)}
{f_{\vX}^{eq}(\vX_{i};\alpha_{i-1}) Z(\alpha_{i-1})}
\cdot \prod_{i=1}^N P(\vX_i|\vX_{i-1};\alpha_{i-1}) f_{\vX}^{eq}(\vX_0;\alpha_0)
\nonumber\\
= &\dfrac{Z(\alpha_n)}
{Z(\alpha_{0})}
\int\cdots\int \prod_{i=0}^N \rd \vX_i \
\prod_{i=1}^N  P(\vX_i|\vX_{i-1};\alpha_{i-1})
\prod_{i=1}^N f_{\vX}^{eq}(\vX_{i-1};\alpha_{i-1})
\Bigg/
\prod_{i=1}^N f_{\vX}^{eq}(\vX_{i};\alpha_{i-1}) \nonumber\\
= &\dfrac{Z(\alpha_n)}{Z(\alpha_{0})}.
\end{align}
Since we have defined in Sec. \ref{sec:2.1} the free energy as: $A_2(\alpha) = - \epsilon^2 \ln Z(\alpha)$, thus we obtain the
Jarzynski equality:
\begin{align}
\left\langle e^{-\frac{1}{\epsilon^2} W[\vX(\tau),\alpha(\tau)]}\right\rangle_{\big[\vX(\tau)
,\alpha(\tau)\big]}
= e^{-\frac{1}{\epsilon^2} \Delta A_2}.
\end{align}

In a very similar vein, for the macroscopic thermodynamic
variables $(E,\widetilde\alpha)$, one defines the work done to
the system by the external environment through controlling $\widetilde\alpha(t)$ with rate $\dot{\widetilde\alpha}$:
\begin{align}
W\big[E(\tau),\widetilde\alpha(\tau)\big]
= - \int_0^t \widetilde{F}_{\widetilde\alpha}(E,\widetilde\alpha) \dot{\widetilde\alpha} \rd \tau.
\end{align}
Write $\varphi_\tau(\widetilde\alpha) = E(\tau,\widetilde\alpha)$.
Then the discretized
$\left\langle e^{-\frac{1}{\epsilon^2} W[E(\tau),\widetilde\alpha(\tau)]}\right\rangle_{\big[E(\tau),\widetilde\alpha(\tau)\big]}$ is:
\begin{align}
&\left\langle e^{-\frac{1}{\epsilon^2} W[E_0,\cdots,E_N ; \widetilde\alpha_0,\cdots,\widetilde\alpha_N]}\right\rangle_{\big[E_0,\cdots,E_N ; \widetilde\alpha_0,\cdots,\widetilde\alpha_N\big]} \nonumber\\
= &\int\cdots\int \prod_{i=0}^N \rd E_i
 \prod_{i=1}^N P(E_i|E_{i-1};\widetilde\alpha_{i-1}) p(E_0,t_0;\widetilde\alpha_0) \times
\nonumber\\
& \exp\left(-\sum_{i=1}^N
 \frac{\varphi_i(\widetilde\alpha_i)- \varphi_i(\widetilde\alpha_{i-1})} {\epsilon^2}
+ S(E_i,\widetilde\alpha_i) - S(E_i,\widetilde\alpha_{i-1}) \right),
\end{align}
where $S(E,\widetilde{\alpha})$ is defined in Eq. (\ref{S_entropy}).
On the other hand, equilibrium probability density function of $E$ at $(E_i,\widetilde\alpha_i)$ is:
\begin{align}
f^{eq}_{\mathbf{E}}(E_i,\widetilde\alpha_i)
&= \dfrac{1}{Z(\widetilde\alpha_i)} \oint_{\varphi(\vx;\widetilde\alpha) = E_i} e^{-\varphi(\vx;\widetilde\alpha_i)/\epsilon^2} \dfrac{\rd \Sigma^{n-1}}
{||\nabla_\vx \varphi(\vx;\widetilde\alpha_i)||} \nonumber\\
&= Z^{-1}(\widetilde\alpha_i)
        \exp\left(-\frac{\varphi_i(\widetilde\alpha_i)}{\epsilon^2} + S(E_i,\widetilde\alpha_i)\right).
\end{align}
Hence, we have
\begin{align}
&\left\langle e^{-\frac{1}{\epsilon^2} W[E_0,\cdots,E_N ; \widetilde\alpha_0,\cdots,\widetilde\alpha_N]}\right\rangle_{\big[E_0,\cdots,E_N ; \widetilde\alpha_0,\cdots,\widetilde\alpha_N\big]} \nonumber\\
= &\int\cdots\int \prod_{i=0}^N \rd E_i \ \prod_{i=1}^N
\dfrac{f_{\mathbf{E}}^{eq}(E_i,\widetilde\alpha_i) Z(\widetilde\alpha_i)}
{f_{\mathbf{E}}^{eq}(E_{i},\widetilde\alpha_{i-1}) Z(\widetilde\alpha_{i-1})}
\left( \prod_{i=1}^N P(E_i|E_{i-1},\widetilde\alpha_{i-1}) f_{\mathbf{E}}^{eq}(E_0,\widetilde\alpha_0) \right)
\nonumber\\
= &\dfrac{Z(\widetilde\alpha_n)}{Z(\widetilde\alpha_{0})}
\int\cdots\int \prod_{i=0}^N \rd E_i \
\prod_{i=1}^N  P(E_i|E_{i-1},\widetilde\alpha_{i-1})
\prod_{i=1}^N f_{\mathbf{E}}^{eq}(E_{i-1},\widetilde\alpha_{i-1})
\Bigg/
\prod_{i=1}^N f_{\mathbf{E}}^{eq}(E_{i},\widetilde\alpha_{i-1}) \nonumber\\
= &\dfrac{Z(\widetilde\alpha_n)}{Z(\widetilde\alpha_{0})}
= e^{-\frac{1}{\epsilon^2}\Delta A_2(\alpha)}.
\end{align}
Therefore, the log-mean exponential of minus work is equal to the minus of free energy difference.

\subsection{Crooks' approach}

	Instead of introducing stochastic work functional, G. E. Crooks' approach recognizes the important role of {\em time reversal}
trajectory $\check{\vX}(t)$, and the deep relationship between work, energy, and dissipation, e.g., entropy production.  Let us consider the path probability of a backward trajectory $\mathcal{P}[\check{\vX}(\tau)|\check{\vX}(0);\check{\alpha}(\tau)]$ against a forward one $\mathcal{P}[\vX(\tau)|\vX(0);\alpha(\tau)]$, in which $\left(\check{\vX}(\tau);\check{\alpha}(\tau) \right)
= \left( \vX(t-\tau);\alpha(t-\tau) \right)$, where the initial and final
distribution of $\vX(\tau)$ follow the equilibrium distribution.
We solve for $\mathcal{P}[\vX(\tau)|\vX(0);\alpha(\tau)]$
from the probability of a Brownian motion whose increments
are multivariate Gaussian:
\begin{equation}
\frac{1}{\epsilon}\Gamma(\alpha_i)^{-1}\Big(
         \vX_{i+1}-\vX_{i}
  - M(\alpha_i) \vX_i \Delta \tau \Big)
             =  \vB_{t_{i+1}}-\vB_{t_i}.
\end{equation}

Hence, the probability density functional of a path $\big[\vX_0,\cdots,\vX_N | \vX_0 ; \alpha_0,\cdots,\alpha_N\big]$ is:
 \begin{align}
&\mathcal{P}\big[\vX_0,\cdots,\vX_N | \vX_0 ; \alpha_0,\cdots,\alpha_N\big]
= \prod_{i=0}^{N-1} P\big(\vX_{i+1}|\vX_i;\alpha_i\big)
\nonumber\\
= &\prod_{i=0}^{N-1} \dfrac{1}{(\pi\Delta \tau)^{n/2}}
e^{-\frac{1}{\epsilon^2\Delta \tau} \left(\Gamma(\alpha_i)^{-1}(\vX_{i+1}-\vX_{i})
- \Gamma(\alpha_i)^{-1} M(\alpha_i) \vX_i \Delta \tau\right)^2
 }.
\end{align}
Here $\big(\mathbf{v}(\vX,\alpha)\big)^2\equiv \big(\mathbf{v}(\vX,\alpha)\big)^T\big(\mathbf{v}(\vX,\alpha)\big)$.
Then
the probability density functional of the inverse path $\big[\vX_N,\cdots,\vX_0 | \vX_N ; \alpha_N,\cdots,\alpha_0\big]$ is:
\begin{align}
&\mathcal{P}\big[\vX_N,\cdots,\vX_0 | \vX_N ; \alpha_N,\cdots,\alpha_0\big]
= \prod_{i=1}^{N} P\big(\vX_{i-1}|\vX_{i};\alpha_i\big)
\nonumber\\
= &\prod_{i=1}^{N} \dfrac{1}{(\pi\Delta \tau)^{n/2}}
e^{-\frac{1}{\epsilon^2 \Delta \tau}\left(\Gamma(\alpha_i)^{-1}(\vX_{i-1}-\vX_{i})
- \Gamma(\alpha_i)^{-1} M(\alpha_i) \vX_i \Delta \tau\right)^2
 }.
\end{align}
Therefore, offsetting by a normalization factor, an infinite-dimensional
functional integral,
\begin{align}
\mathcal{P}\big[\vX(\tau)|\vX(0);\alpha(\tau)\big]
&\propto \exp \left[
-\frac{1}{\epsilon^2} \int_0^t \left(\Gamma(\alpha)^{-1} \rd \vX/\sqrt{\rd \tau} - \Gamma(\alpha)^{-1} M(\alpha) \vX \sqrt{\rd \tau}\right)^2 \right]
\nonumber\\
&= \exp\left[ -\frac{1}{\epsilon^2} \int_0^t \left(\Gamma(\alpha)^{-1}\dot{\vX} - \Gamma(\alpha)^{-1} M(\alpha) \vX\right)^2 \rd \tau \right].
\end{align}
Probability of the backward path can be found by substituting $\tau$
with $t-\tau$:
\begin{align}
\mathcal{P}\big[\check{\vX}(\tau)|\check{\vX}(0);\check{\alpha}(\tau)\big]
\propto \exp\left[ -\frac{1}{\epsilon^2}
\int_0^t \left(-\Gamma(\alpha)^{-1}\dot{\vX} - \Gamma(\alpha)^{-1} M(\alpha) \vX\right)^2 \rd \tau \right].
\end{align}
Therefore, we have an equality for heat dissipation:
\begin{align}
&\dfrac{\mathcal{P}[\check{\vX}(\tau)|\check{\vX}(0);\check{\alpha}(\tau)]}
{\mathcal{P}[{\vX}(\tau)|{\vX}(0);{\alpha}(\tau)]}
\nonumber\\
&= \exp\left[ \frac{2}{\epsilon^2} \int_0^t
\left(\dot{\vX}^T (\Gamma(\alpha)\Gamma(\alpha)^T)^{-1} M(\alpha) \vX
+ \vX^T M(\alpha)^T (\Gamma(\alpha)\Gamma(\alpha)^T)^{-1} \dot{\vX}\right) \ \rd \tau \right]
\nonumber\\
&= \exp\left[ \frac{2}{\epsilon^2} \int_0^t
\left(\dot{\vX}^T D^{-1}(\alpha)M(\alpha) \vX\right) \ \rd \tau \right] \nonumber\\
&= \exp\left[ \frac{2}{\epsilon^2} \int_0^t \dot{\vX}^T \Xi^{-1}(\alpha) \vX \ \rd \tau
+ \frac{2}{\epsilon^2} \int_0^t \dot{\vX}^T
\big(D^{-1}(\alpha) M(\alpha) - \Xi^{-1}(\alpha)\big) \vX \ \rd \tau \right]
\nonumber\\
&= \exp\left\{ \frac{Q\big[\vX(\tau),\alpha(\tau)\big]}{\epsilon^2}
+ \frac{Q_{hk}\big[\vX(\tau),\alpha(\tau)\big]}{\epsilon^2}  \right\}.
\end{align}
Hatano and Sasa, following Oono and Paniconi,
called the term
\begin{equation}
Q_{hk}\big[\vX(\tau),\alpha(\tau)\big]
=-2\int_0^t \dot{\vX}^T
\big(D^{-1}(\alpha) M(\alpha) - \Xi^{-1}(\alpha)\big) \vX \rd \tau
\end{equation}
house-keeping heat  \cite{sasa}.

Since we start and end with equilibrium distributions with
the corresponding $\widetilde{\alpha}$,
\begin{align}
p\big(\vX_0;\alpha_0\big) &=f_{\vX}^{eq}(\vX_0;\alpha_0) = \dfrac{1}{Z(\alpha_0)} \exp\left[
- \frac{\vX_0^T U(\alpha_0) \vX_0}{\epsilon^2}  \right];
\nonumber\\
p\big(\vX_N;\alpha_N\big) &=f_{\vX}^{eq}(\vX_N;\alpha_N) = \dfrac{1}{Z(\alpha_N)} \exp\left[
- \frac{\vX_N^T U(\alpha_N) \vX_N}{\epsilon^2}  \right].
\end{align}
Therefore,
\begin{eqnarray}
\dfrac{\mathcal{P}[\check{\vX}(\tau),\check{\alpha}(\tau)]}
{\mathcal{P}[{\vX}(\tau),{\alpha}(\tau)]}
 &=&\dfrac{\mathcal{P}[\check{\vX}(\tau)|\check{\vX}(0);\check{\alpha}(\tau)]p(\vX_N;\alpha_N)}
{\mathcal{P}[{\vX}(\tau)|{\vX}(0);{\alpha}(\tau)]p(\vX_0;\alpha_0)}
\nonumber\\
&=&\exp\left( \frac{ Q[\vX(\tau),\alpha(\tau)]
- Q_{hk}[\vX(\tau),\alpha(\tau)]
- \Delta \varphi + \Delta A_2 }{ \epsilon^2 } \right)
\nonumber\\
&=&\exp\left( \frac{- W[\vX(\tau),\alpha(\tau)]
- Q_{hk}[\vX(\tau),\alpha(\tau)] + \Delta A_2 }{\epsilon^2}\right) .
\end{eqnarray}
Now taking ensemble average of the trajectories $[\vX(\tau),\alpha(\tau)]$ over $\dfrac{\mathcal{P}[\check{\vX}(\tau),\check{\alpha}(\tau)]}{\mathcal{P}[\vX(\tau),\alpha(\tau)]}$ gives:
\begin{eqnarray}
&& \int \mathcal{D} [\vX(\tau),\alpha(\tau)] \
\mathcal{P}[\check{\vX}(\tau),\check{\alpha}(\tau)]
 \ = \  \left\langle\dfrac{\mathcal{P}[\check{\vX}(\tau),\check{\alpha}(\tau)]}
{\mathcal{P}[\vX(\tau),\alpha(\tau)]}\right\rangle_{[\vX(\tau),\alpha(\tau)]} \nonumber\\
&=& e^{ \Delta A_2}\Big\langle e^{ -W[\vX(\tau),\alpha(\tau)] - Q_{hk}[\vX(\tau),\alpha(\tau)]}\Big\rangle_{\big[\vX(\tau),\alpha(\tau)\big]}.
\end{eqnarray}
When the system is in detailed balance, Crooks' approach recovers the Jarzynski equality.  If one chooses the global energy $\varphi$
with zero reference for each own equilibrium, i.e., $\Delta A_2=0$
for all $\alpha$, then it recovers the Hatano-Sasa equality.

\end{document}